\documentclass[letterpaper]{jpconf}
\usepackage{graphicx}
\begin{document}
\title{Recent results on $\phi_3$ at Belle}

\author{Katsumi Senyo (on behalf of the Belle collaboration)}

\address{Nagoya University, Furo-cho, Chikusa-ku, Nagoya, 464-8601 Japan}

\ead{senyo@senyo.net}

\begin{abstract}
We report results of measurement of the unitarity triangle angle $\phi_3$ with $B^+\rightarrow D^{(*)}K^+$ Dalitz plot analysis and related issues with $B^0 \rightarrow D_s^{*+} \pi^-$ and $B^0 \rightarrow D_s^{*-} K^+$ decay processes.   The Dalitz analysis improves accuracy of the angle $\phi_3$ as $78.4^\circ ~^{+10.8^\circ}_{-11.6^\circ}\pm 3.6^\circ\pm 8.9^\circ$ and the branching fractions are found to be $B^0 \rightarrow D_s^{*+} \pi^-(K^-)$ decays are set as $\mathcal{B}(B^0 \rightarrow D_s^{*+} \pi^-) = (1.75 \pm 0.34 (\mathrm{stat}) \pm 0.17(\mathrm{syst}) \pm  0.11(\mathcal{B})) \times 10^{-5}$ and $\mathcal{B}(B^0 \rightarrow D_s^{*-} K^+) = (2.02 \pm 0.33 (\mathrm{stat}) \pm  0.18({\rm syst}) \pm 0.13(\mathcal{B})) \times 10^{-5}$.
\end{abstract}

\section{Introduction}

Determinations of the Kobayashi-Maskawa matrix elements~[1] provide important checks on the consistency of the standard model and ways to search for new physics.  In decades, several methods have been discussed to provide interior angles $\phi_1$, $\phi_2$ and $\phi_3$ of the unitarity triangle.  The $\phi_3$ measurement is still on the way since the measurement has been statistically limited even with modern electron-positron B factories.  Two of those analyses to evaluate $\phi_3$ are reported here.  As space is limited, Belle experiment and KEKB accelerator are not explained here.  Detailed description of the Belle detector is found elsewhere~[2].  

\section{Measurement of the branching fractions for $B^0 \rightarrow D_s^{*+} \pi^-$ and $B^0 \rightarrow D_s^{*-} K^+$}

The time-dependent $CP$ analysis of the $B^0(\bar{B}^0) \rightarrow D^{*\mp}\pi^{\pm}$ system provides a theoretically 
clean measurement of the product $R_{D^*\pi} \sin (2\phi_1 + \phi_3)$~[3], where $R_{D^*\pi}$ is the ratio of the 
magnitudes of the doubly Cabibbo-suppressed decay amplitude to the Cabibbo-favoured decay amplitude.  Unlike the $B^0 \rightarrow D^{* \mp} \pi^\pm$ process, $B^0 \rightarrow D_s^{*+} \pi^-$, which is predominantly a spectator process with a $b \rightarrow u$ transition, does not have contributions from $\bar{B}^0$ decays to the same final state and can provide clean experimental access to $R_{D^* \pi}$.  Assuming SU(3) flavour symmetry between $D^*$ and $D_s^*$, $R_{D^* \pi}$ is given by

\begin{equation}
R_{D^* \pi} = \tan \theta_C (\frac{f_{D^*}}{f_{D_s^*}}) \sqrt{\frac{\mathcal{B}(B^0 \rightarrow D_s^{*+} \pi^-)}{\mathcal{B}(B^0 \rightarrow D^{*-} \pi^+)}},
\end{equation}
where $\theta_C$ is the Cabibbo angle, $f_{D^*}$ and $f_{D_s^*}$ are the meson form factors, and the $\mathcal{B}$'s stand for the corresponding branching fractions.  The $B^0 \rightarrow D_s^{*+} \pi^-$ process, in addition, does not have a penguin loop contribution and hence can in principle be used to determine $|V_{ub}|$~[4].

The decay $B^0 \rightarrow D_s^{*+} \pi^-$ does not have a contribution from the $W$-exchange amplitude, as the quark-antiquark pair with the same flavor, required for such a diagram, is absent from the final state.  We assume the $W$-exchange contributions in $B^0 \rightarrow D^{*\mp} \pi^\pm$ to be negligible, in making the correspondence between $D^{*+}\pi^-$ and $D_s^{*+}\pi^-$ in the $R_{D^* \pi}$ calculation.  The size of the W-exchange diagram can be estimated from the $B^0 \rightarrow D_s^{*-} K^+$ decay, which proceeds only via $W$ exchange.  The $B^0 \rightarrow D_s^{*-} K^+$ branching fraction was expected to be enhanced due to rescattering effects~[5, 6]. 
Here we briefly report an improved measurement of the branching fractions for $B^0 \rightarrow D_s^{*+} \pi^-$ and $B^0 \rightarrow D_s^{*-} K^+$ with a data sample consisting of $657 \times 10^6$ $B\bar{B}$ pairs.  Detailed description of the analysis is found elsewhere~[7].

The signal is reconstructed in three $D_s^+$ modes: $\phi \pi^+$ with $\phi \rightarrow K^+ K^-$, $\bar{K}^*(892)^0K^+$ with $\bar{K}^*(892)^0 \rightarrow K^- \pi^+$, and $K_S^0 K^+$ with $K_S^0 \rightarrow \pi^+\pi^-$.  
After applying Belle standard event selection criteria, we obtained results of fits~(Figure 1) summarized in Table 1. 
The significance is defined as $\sqrt{-2 \ln (\mathcal{L}_0/\mathcal{L}_{max})}$, where $\mathcal{L}_{max}(\mathcal{L}_0)$ are the likelihoods for the best fit and with the signal branching fraction fixed to zero.  Table 2 summarizes the systematic uncertainties involved.  The overall systematic uncertainty is obtained by adding these contributions in quadrature.
\begin{table}
\caption{Efficiency ($\epsilon$), yield($N_\mathrm{sig}$), branching fraction ($\mathcal{B}$), and statistical
 significance not including systematic uncertainties ($\mathcal{S}$) from the fits to the data obtained individually in the three
 $D_s^+$ modes as well as from the simultaneous fit.  The second uncertainty on the $\mathcal{B}$'s is due to the uncertainties
 in $D_s^+$ decay branching fractions.}
\begin{center}
\begin{tabular}{lcrllr}
\br
$B^0$ mode & $D_s^+$ mode & $\epsilon$(\%) & $N_{\rm sig}$ & $\mathcal{B}(10^{-5})$ & $\mathcal{S}(\sigma)$ \\
\mr
$B^0 \rightarrow D_s^{*+} \pi^-$ & $\phi(K^+K^-)\pi^+$ & $15.2$ & $32 \pm 8$ & $1.58\pm 0.40\pm 0.24$ & $3.2$ \\
& $\bar{K}^+(892)^0(K^-\pi^+)K^+$ & $7.9$ & $29 \pm 10$ & $2.30 \pm 0.60 \pm 0.35 $ & $2.6$ \\
& $K_S^0K^+$ & $8.0 $ & $ 13 \pm 7$ & $1.78 \pm 0.92 \pm 0.11$ & $2.2$ \\
& Simultaneous  & $\cdots$ & $\cdots$ & $1.75 \pm 0.34 \pm 0.11$ & $6.6$ \\
$B^0 \rightarrow D_s^{*-} K^+$ & $\phi(K^+K^-)\pi^+$ & $13.4$ & $33 \pm 8$ & $1.81 \pm 0.41\pm 0.27$ & $3.2$ \\
& $\bar{K}^+(892)^0(K^-\pi^+)K^+$ & $6.4 $ & $23 \pm 7$ & $2.22 \pm 0.66 \pm 0.34 $ & $2.8$ \\
& $K_S^0K^+$ &  $6.9$ & $14 \pm 5 $ & $ 2.14 \pm 0.80 \pm 0.13 $& $3.1$ \\
& Simultaneous  & $\cdots$ & $\cdots$ & $2.02 \pm 0.33 \pm 0.13$ & $8.6$ \\
\br
\end{tabular}
\end{center}
\end{table}
\begin{figure}
\begin{minipage}{.38\textwidth}
\begin{center}
\includegraphics[scale=.65,clip]{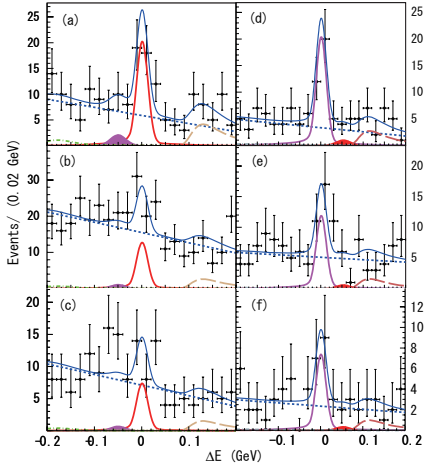}
\caption{The simultaneous fit in the $B^0 \rightarrow D_s^{*+} \pi^-$ [(a)$\phi\pi$, (b)$K^{*0}K$, and (c)$K_S^0K$ modes] and the $B^0 \rightarrow D_s^{*-} K^+$[(d) - (f)] signal modes.  Signal peaks are shown by the solid curves, while the solid-filled curves represent the cross-feed.}
\end{center}
\end{minipage}\hspace{0.02\textwidth}%
\begin{minipage}{.6\textwidth}
\begin{center}
\makeatletter
\def\@captype{table}
\makeatother
\caption{Contributions to the systematic uncertainty.}
\begin{tabular}{lcc}
\br
Source & \multicolumn{2}{c}{Contribution (\%)} \\
\mr
& $D_s^{*+} \pi^-$ &  $D_s^{*+} K^-$ \\
$D_s^+$ branching fraction uncertainties & & \\
Signal & 5.9 & 6.2 \\
Peaking background & 1.5 & 1.9 \\
Total($\mathcal{B}$) & 6.1 & 6.5 \\
\mr
Tracking efficiency & 4.0 &4.0 \\
Photon detection efficiency & 7.0 & 7.0 \\
Particle identification efficiency & 2.4 & 2.1 \\
$K_S^0$ efficiency & 1.1 & 1.1 \\
$\mathcal{LR}$ & 0.6 & 0.5 \\
$N_{B\bar{B}}$ & 1.4 & 1.4 \\
MC statistics & 1.4 & 1.6 \\
PDF shape & 3.4 & 1.5 \\
Fit bias & 0.9 & 0.3 \\
Total (other) & 9.4 & 8.8\\
\br
\end{tabular}
\end{center}
\end{minipage}
\end{figure}

We obtain $\mathcal{B}(B^0 \rightarrow D_s^{*+} \pi^-) = (1.75 \pm 0.34 (\mathrm{stat}) \pm 0.17(\mathrm{syst}) \pm
 0.11(\mathcal{B})) \times 10^{-5}$ and $\mathcal{B}(B^0 \rightarrow D_s^{*-} K^+) = (2.02 \pm 0.33 (\mathrm{stat}) \pm
 0.18({\rm syst}) \pm 0.13(\mathcal{B})) \times 10^{-5}$ with significance of $6.1 \sigma$ and $8.0 \sigma$, respectively,
 where the systematic uncertainties on the signal yield as well as the statistical uncertainties are included in the
 significance evaluation.  Using observed value for the $B^0 \rightarrow D_s^{*+} \pi^-$ branching fraction, the latest
 values for $\mathcal{B}(B^0 \rightarrow D^{*-} \pi^+) = (2.76 \pm 0.13) \times 10^{-3}$, $\tan \theta_C = 0.2314 \pm 0.0021$~[8], and the theoretical estimate of the ratio $f_{D_s^+}/f_{D^+} = (1.164 \pm 0.006({\rm stat}) \pm 0.020({\rm syst}))$~[9], we obtain
 \begin{equation}
 R_{D^*\pi} = (1.58 \pm 0.15({\rm stat}) \pm 0.10 ({\rm syst}) \pm 0.03({\rm th})) \%,
 \end{equation}
where the first error is statistical, the second corresponds to the experimental systematic uncertainty, and the third
 accounts for the theoretical uncertainty in the $f_{D_s^+}/f_{D^+}$ estimation.  We have assumed that the ratio
 $f_{D_s}/f_{D}$ is equal to the ratio of vector meson decay constants, $f_{D_s^*}/f_{D^*}$.  The value we obtain for $R_{D^*\pi}$ is
 consistent with theoretical expectation of 2\%.
 
 The observed value for the $B^0 \rightarrow D_s^{*-} K^+$ branching fraction is 2 orders of magnitude lower than that for the Cabibbo-favoured decay $B^0 \rightarrow D^{*-} \pi^+$.

\section{$\phi_3$ measurement with a Dalitz plot analysis of $B^+ \rightarrow D^{(*)} K^+$ decay}
Among various methods to measure the angle  $\phi_3$ using $CP$ violation in $B\rightarrow DK$ decays~[10-14], three body final states such as $K_S^0 \pi^+ \pi^-$ have been suggested as promising modes for the extraction of 
$\phi_3$~[15].  Assuming no $CP$ asymmetry in neutral $D$ decays, the amplitude of the neutral $D$ decays from $B^\pm 
\rightarrow DK^\pm$ as a function of Dalitz plot variables $m_+^2 = m_{K_S^0\pi^+}^2$ and $m_-^2 = m_{K_S^0\pi^-}^2$ is 
\begin{equation}
M_\pm = f(m_\pm^2, m_\mp^2) + r e^{\pm i\phi_3 + i\delta} f(m_\mp^2, m_\pm^2),
\end{equation}
where $f(m_+^2, m_-^2)$ is the amplitude of the $\bar{D}^0 \rightarrow K_S^0\pi^+\pi^-$ decay, $r$ is the ratio of the magnitudes of the two interfering amplitudes, and $\delta$ is the strong phase difference between them.  The $\bar{D}^0 \rightarrow K_S^0\pi^+\pi^-$ decay amplitude $f$ can be determined from a lage sample of flavor-tagged $\bar{D}^0 \rightarrow K_S^0\pi^+\pi^-$ decays produced in $e^+e^- \rightarrow q\bar{q}$ continuum process.  Once $f$ is known, a simultaneous fit of $B^+$ and $B^-$ data allows the contribution of $r$, $\phi_3$ and $\delta$ to be separated.  

In this paper, we report a preliminary result of a measurement of $\phi_3$ using the modes $B^+ \rightarrow DK^+$ and $B^+ \rightarrow D^*K^+$ with $\bar{D}^0 \rightarrow K_S^0 \pi^+ \pi^-$, based on a 605 ${\rm fb}^{-1}$ data sample.  Detailed description is found in reference~[16].
After event selection(Figure~2 and 3), we obtained $\phi_3$ and other parameters by fitting the Dalitz plots for two event samples, $B^+ \rightarrow DK^+$, $B^+ \rightarrow D^*K^+$ and combined results of angle $\phi_3$ as $78.4^\circ ~^{+10.8^\circ}_{-11.6^\circ}\pm 3.6^\circ\pm 8.9^\circ$~(Table~3).   
\begin{table}
\begin{center}
\caption{$CP$ fit results. The first error is statistical, the second is experimental systematic, and the third is the model uncertainty.  The error in combined result of $\phi_3$ is only statistical. Consult the reference for detailed results~[16].}
\begin{tabular}{|l|c|c|c|c|}
\hline
 Parameter & $1\sigma$ interval & $2\sigma$ interval & Systematic error & Model uncertainty \\
\hline
$\phi_3$ & $(78.4^{+10.8}_{-11.6})^\circ$ & $54.2^\circ < \phi_3 < 100.5^\circ $ & $3.6^\circ$ & $8.9^\circ$ \\
$r_{DK}$ & $0.160^{+0.040}_{-0.038}$ & $0.084 < r_{DK} < 0.239 $ & $0.011$ & $^{+0.050}_{-0.010}$ \\
$r_{D^*K}$ & $0.196^{+0.072}_{-0.069}$ & $0.061 < r_{D^*K} < 0.271 $ & $0.012$ & $^{+0.062}_{-0.012}$ \\
$\delta_{DK}$ & $(136.7^{+13.0}_{-15.8})^\circ$ & $102.2^\circ < \delta_{DK} < 162.3^\circ$ & $4.0^\circ$ & $22.9^\circ$ \\
$\delta_{D^*K}$ & $(341.9^{+18.0}_{-19.6})^\circ$ & $296.5^\circ < \delta_{D^*K} < 382.7^\circ$ & $3.0^\circ$ & $ 22.9^\circ$ \\
\hline
\end{tabular}
\end{center}
\end{table}
\begin{figure}
\begin{minipage}{.48\textwidth}
\begin{center}
\includegraphics[scale=1.0,clip]{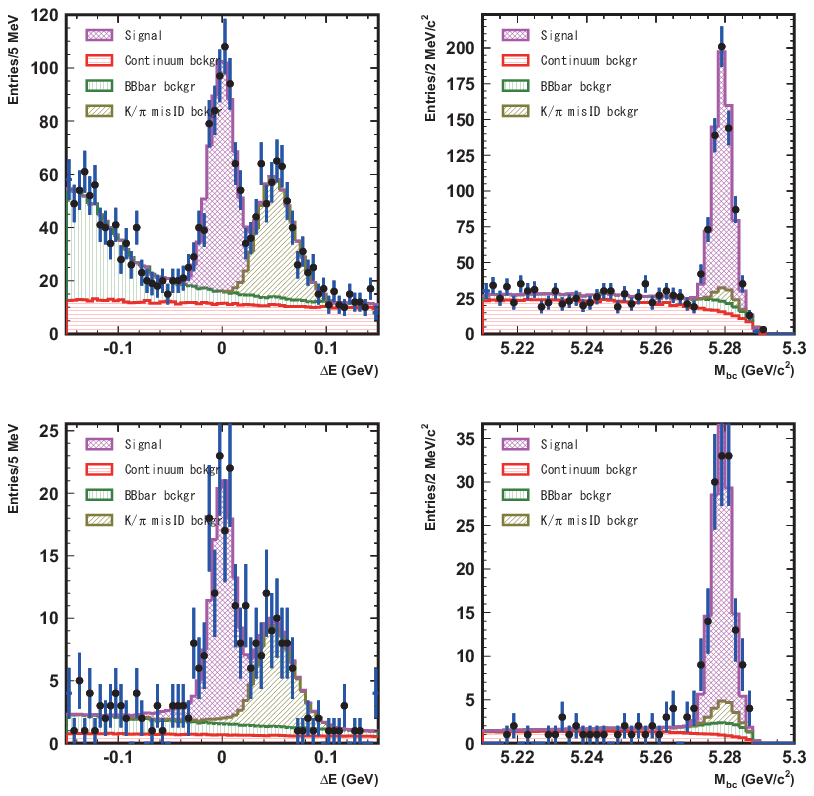}
\caption{$\Delta E$ and $M_{bc}$ distributions for the $B^+\rightarrow DK^+$(top) and $B^+\rightarrow D^*K^+$(bottom) event samples.  Points with error bars are the data and the histogram is the result of a MC simulation according to the fit result.}
\end{center}
\end{minipage}\hspace{0.04\textwidth}%
\begin{minipage}{.48\textwidth}
\begin{center}
\includegraphics[scale=1.0,clip]{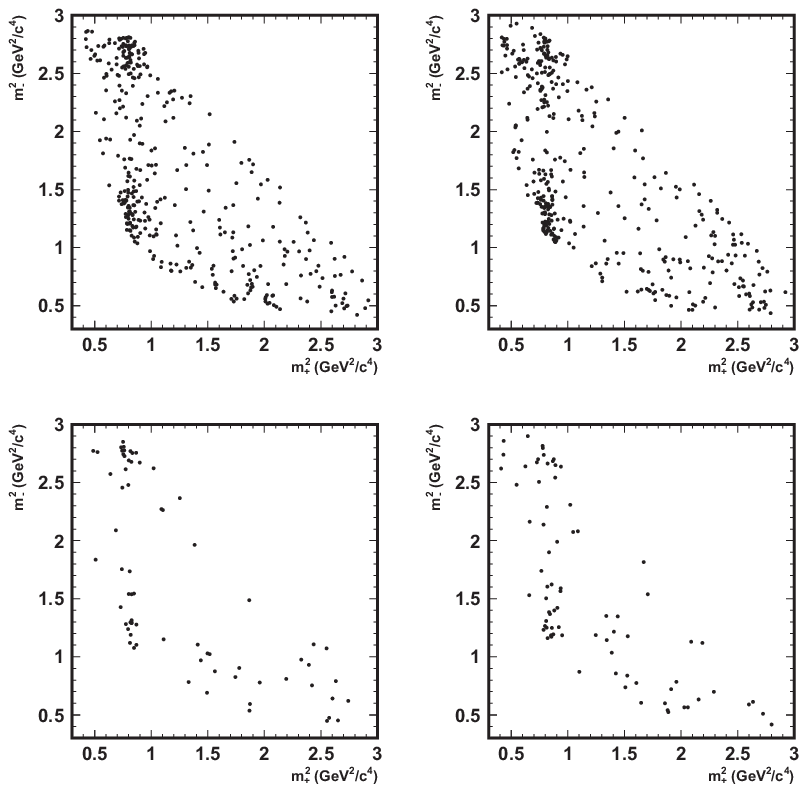}
\caption{Dalitz distributions of $\bar{D}^0 \rightarrow K_S^0\pi^+\pi^-$ decays from selected $B^+\rightarrow DK^+$(top) and $B^+\rightarrow D^*K^+$(bottom) candidates, shown separately for $B^-$(left) and $B^+$(right) tags.}
\end{center}
\end{minipage}
\end{figure}
\section{Conclusion}
We report results of measurements of the angle $\phi_3$ of Kobayashi-Maskawa triangle with $B^+\rightarrow D^{(*)}K^+$ Dalitz plot analysis and related issues with $B^0 \rightarrow D_s^{*+} \pi^-$ and $B^0 \rightarrow D_s^{*-} K^+$ decay processes.  Measurement of angle $\phi_3$ is on the way to verify the Standard Model prediction, and will be a rich physics subject for a decade with next generation $B$ factories and LHC.

\end{document}